\documentclass{svmult}

\usepackage{mathptmx}
\usepackage{helvet}
\usepackage{courier}
\usepackage{graphicx}
\usepackage{makeidx}
\usepackage{multicol}
\usepackage{footmisc}
\usepackage{subeqnarray}

\newcommand\dnot{{\hat\partial_0}}

\newcommand\be{\begin{equation}}
\newcommand\bea{\begin{subeqnarray}}
\newcommand\ee{\end{equation}}
\newcommand\eea{\end{subeqnarray}}
\newcommand\plusequal{\lower 2pt \hbox{$\, \buildrel {+} \over {=} \,$}}

\begin{document}
  
\title*{Strongly Hyperbolic Extensions of the ADM Hamiltonian}
\author{J.~David Brown}
\institute{Department of Physics, North Carolina State University,
Raleigh, NC 27695 USA}
\maketitle

\abstract{The ADM Hamiltonian formulation of general relativity with prescribed lapse 
and shift is a weakly hyperbolic system of partial differential equations.
In general weakly hyperbolic systems are not mathematically 
well posed. For well posedness, the theory should be reformulated so that the complete 
system, evolution equations plus gauge conditions, is (at least) strongly hyperbolic. 
Traditionally, reformulation has been carried out at the level of equations of 
motion. This typically destroys the variational and Hamiltonian structures 
of the theory. Here I show that one can extend the ADM formalism to (i) 
incorporate the gauge conditions as dynamical equations and (ii) affect the 
hyperbolicity of the complete system, all while maintaining a Hamiltonian 
description.  The extended ADM formulation is used to obtain a strongly 
hyperbolic Hamiltonian description of Einstein's theory that is 
generally covariant under spatial diffeomorphisms and 
time reparametrizations, and has physical characteristics. 
The extended Hamiltonian formulation with 1+log slicing and gamma--driver 
shift conditions is weakly hyperbolic.}

\section{Introduction}
This paper is dedicated to Claudio Bunster in celebration of his sixtieth birthday.
In a remarkable body of work Claudio showed that we can view the Hamiltonian 
formulation of general relativity as fundamental. 
(See in particular references \cite{Teitelboim:1972vw,Hojman:1976vp,Claudio:Held}.) 
He considered the requirement that the 
sequence of spatial three--geometries evolved by the Hamiltonian should be 
interpretable as a four--dimensional spacetime. 
From this assertion and a few modest assumptions he was able to derive the ADM Hamiltonian
\cite{Dirac:1958sc,ADM:Witten} of general relativity. A number of 
deep insights into the nature of gravity and matter came from his analysis, including the role  
of gauge symmetries in electrodynamics and Yang--Mills theories and the necessity for all matter fields 
to couple to gravity. 

In general relativity we are faced with the practical problem of predicting the 
future evolution of strongly gravitating systems, including interacting black holes and neutron stars. 
Such problems fall into the realm of numerical relativity. 
Naturally, the first attempts at numerical modeling 
in general relativity were based on the ADM Hamiltonian 
equations. By the early 1990's it became clear that the ADM equations 
were not appropriate for numerical computation because they are not well posed in a mathematical sense. 
What followed was more than a decade of activity in which the ADM equations were rewritten in a variety of ways.
The goal was to produce a well posed system of partial differential equations (PDE's) for Einstein's theory. 
One strategy for modifying the ADM equations was to add multiples of the constraints to the right--hand sides.
Another strategy was to introduce new independent variables defined as combinations of metric 
tensor components and their spatial derivatives. This later strategy introduces new constraints into the system, 
namely, the constraints that the definitions of the new variables should hold for all time. 

In a practical sense, the effort to re--express the ADM equations has been successful. Currently there 
are a number of  formulations of the Einstein evolution equations that appear to be ``good enough", the most 
popular for numerical work being the BSSN system.\cite{Shibata:1995we,Baumgarte:1998te}
BSSN relies on a conformal splitting of the metric and extrinsic curvature. It introduces new independent 
variables, the ``conformal connection functions", defined as the trace (in its lower indices) of the Christoffel 
symbols built from the conformal metric. 

BSSN and the other ``modern" formulations of Einstein's theory are very clever. But 
at a basic level, these formulations are obtained through a manipulation of the equations of motion as a 
system of PDE's. What is invariably lost is the beautiful Hamiltonian structure found in the ADM 
formulation. In this paper I present a systematic procedure that can be used to modify 
the ADM equations in an effort to obtain a good system of PDE's without losing the Hamiltonian structure. 

A good system of PDE's is one that is mathematically well posed. As a general rule, a system 
formulated in space without boundaries must be strongly hyperbolic to be well posed. If boundaries 
are present, an even stronger notion of hyperbolicity, symmetric hyperbolicity, is needed to prove 
well posedness. We are interested in extensions of the ADM equations that, like the ADM equations themselves, have 
first--order time and second--order space derivatives. It turns out that a simple prescription can 
be given to test for strong hyperbolicity in such systems of PDE's. 
The justification 
for this prescription requires a rather deep mathematical analysis, but the prescription itself is fairly easy 
to apply. In section II, I discuss hyperbolicity and justify the prescription for well posedness with 
heuristic arguments. 

Another issue that has become apparent from recent numerical work is the benefit, in practice, of  
incorporating the slicing and coordinate conditions (gauge conditions) as dynamical equations. That is, the 
lapse function and shift vector are not fixed a priori but are determined along with the 
other fields through evolution equations of their own. The hyperbolicity of the entire system 
of PDE's, including the equations for the lapse and shift, must be considered. The issues of 
gauge conditions and hyperbolicity cannot be separated. 

In this paper I show that the Hamiltonian formulation of general relativity can be 
extended to (i) incorporate dynamical gauge conditions and (ii) alter the 
level of hyperbolicity. In section III the ADM formulation is enlarged by the introduction of 
momentum variables $\pi$ and $\rho_a$ conjugate to the lapse function $\alpha$ and shift 
vector $\beta^a$. In this way the lapse and shift become dynamical. 
The new momenta are primary constraints and they appear in the action with 
undetermined multipliers $\Lambda$ and $\Omega^a$. The usual Hamiltonian and momentum constraints, 
${\cal H}$ and ${\cal M}_a$, are secondary constraints. This Hamiltonian formulation of Einstein's 
theory is not new \cite{DeWitt:1967yk,HRT}, and is not substantially different from the original ADM 
formulation---like the ADM formulation, 
it is only weakly hyperbolic. This is shown in section IV. 

The Hamiltonian formulation with dynamical lapse and shift is extended in section V by allowing the 
multipliers $\Lambda$ and $\Omega^a$ to depend on the canonical variables. This has two effects. 
First, it changes the evolution equations for the lapse and shift, yielding gauge conditions that 
depend on the dynamical variables. Second, it changes the principal parts of the evolution equations and 
potentially changes the level of hyperbolicity of the system. The hyperbolicity of the 
extended Hamiltonian formulation is considered in section VI for a fairly general choice of 
multipliers that preserves spatial diffeomorphism invariance and time reparametrization 
invariance. When the multipliers are chosen so that the evolution equations are strongly hyperbolic 
with physical characteristics, the system is equivalent in its principal parts to the generalized 
harmonic formulation of gravity.\cite{Friedrich:1985,Pretorius:2004jg,Lindblom:2005qh} 
It is also shown that the extended Hamiltonian formulation
with 1+log slicing and the gamma--driver shift condition is only weakly hyperbolic. A few final 
remarks are presented in section VII. 

\section{Strong Hyperbolicity for Quasilinear Hamiltonian systems}
Let $q_\mu$, $p_\mu$ denote pairs of canonically conjugate fields. We will consider Hamiltonian
systems for which Hamilton's equations are a quasilinear system of partial differential equations (PDE's). 
Thus we assume that the Hamiltonian $H$ is 
a linear combination of terms that are at most quadratic in the momenta  
and spatial derivatives of the coordinates. More precisely,  $H$ 
should be a linear combination of terms 
$p_\mu p_\nu$, $(\partial_a q_\mu)( \partial_b q_\nu)$, $p_\mu (\partial_a q_\nu)$, 
$p_\mu$, $(\partial_a q_\mu)$, and $1$ with coefficients that depend on the $q$'s.\footnote{Note
that terms proportional to $\partial_a p_\mu$ are also allowed in $H$ since they are 
related to terms of the form $p_\mu (\partial_a q_\nu)$ through integration by parts.}
(Here, $\partial_a$ denotes the derivative with respect to the spatial coordinates.) 
One would like to show that Hamilton's equations are well posed as a system of PDE's. 
The subject of well posedness is a large, active area of research in mathematics and physics.  In this 
section I present a very pedestrian account of the subject in the context of Hamiltonian 
field theory. Much more rigorous and complete discussions can be found 
elsewhere. (See, for example, references 
\cite{Gustafsson,Reula:1998ty,Stewart,Calabrese:2002ej,Reula:2004xd,Nagy:2004td,Gundlach:2005ta,Kreiss:2006mi}.)

A well posed system is one whose solutions 
depend continuously on the initial data. For a well posed system, two sets of initial data that 
are close to one another will evolve into solutions that remain close for some finite 
time. A system is not well posed if it supports modes whose growth rates increase without bound 
with increasing wave number. A concrete example is given below. 

In analyzing well posedness we are primary concerned with the evolution in time of high wave number 
(short wavelength) perturbations of the initial data. For this purpose we can approximate the 
quasilinear system of PDE's by linearizing about a solution. That is, we expand the Hamiltonian to 
quadratic order in perturbations, which we denote $\delta q_\mu$, $\delta p_\mu$. 
We then look for Fourier 
modes of the form $\delta q_\mu = \bar q_\mu e^{i\omega t + i k_a x^a}/(i|k|) $, 
$\delta p_\mu = \bar p_\mu e^{i\omega t + i k_a x^a}$ with nonzero wave number $k_a$. 
Here, $|k| \equiv \sqrt{h^{ab} k_a k_b}$ is the norm of $k_a$ defined in terms of a 
convenient metric $h^{ab}$ (which could be the inverse of the physical spatial metric).
If the ansatz for $\delta q_\mu$, $\delta p_\mu$ is substituted 
into the linearized Hamilton's equations, the system becomes
\be \label{eigenvalueproblem}
	\omega v = (|k|A - iB - C/|k|)v
\ee
where $v$ is the column vector  $v = (\bar q_1,\bar q_2,\ldots, \bar p_1, \bar p_2,\ldots)^T$ of Fourier coefficients. 
Equation (\ref{eigenvalueproblem}) shows that the problem of 
finding perturbative modes with wave number $k_a$ is equivalent to the eigenvalue problem for the matrix $(|k|A - iB - C/|k|)$.
The eigenvector is $v$ and the eigenvalue is the frequency $\omega$. 

What one is really doing in the construction above is replacing the system of PDE's with a 
pseudo--differential system. The factor of $i|k|$ in the denominator of $\delta q_\mu$ is, in effect, equivalent 
to a change of variables in which $q_\mu$ is replaced by $q_\mu/\sqrt{h^{ab}\partial_a \partial_b}$. In this 
way the second order (in space derivatives) PDE's are replaced with an equivalent first order pseudo--differential 
system.\cite{Nagy:2004td}

The behavior of $\omega$ as $|k|$ becomes large depends on the leading order term $A$ in the matrix  $(|k|A - iB - C/|k|)$.
The term $A$ is the ``principal symbol" of the system. It is constructed from the  
coefficients of the highest ``weight" terms in the Hamiltonian, namely, 
the terms proportional to $p_\mu p_\nu$, $(\partial_a q_\mu)( \partial_b q_\nu)$, $p_\mu (\partial_a q_\nu)$ and 
$(\partial_a p_\mu)$. 
Note that it is not necessary to linearize the equations of motion (or expand the Hamiltonian to quadratic order) 
in order to find $A$. In practice we don't actually linearize, we simply identify the coefficients of the 
highest weight terms in the PDE's to form the matrix $A$. 

If $A$ has real eigenvalues and a complete set of eigenvectors that have smooth dependence on the unit vector 
$n_a \equiv k_a/|k|$, the system is said to be {\it strongly hyperbolic}. 
If $A$ has real eigenvalues but the eigenvectors are not complete, the system is said to be {\it weakly hyperbolic}. 
It can be proved that a strongly hyperbolic system of quasilinear PDE's is well--posed.\cite{Nagy:2004td}

Here is the rough idea. Let $S$ denote the matrix whose rows are the left eigenvectors 
of $A$. Assuming strong hyperbolicity, 
the eigenvectors are complete and $S^{-1}$ exists. 
The eigenvalue problem (\ref{eigenvalueproblem}) can be written as  
$\omega \hat v = (|k| \hat A - i\hat B - \hat C/|k|)\hat v$ where 
$\hat v \equiv Sv$, $\hat A\equiv SAS^{-1}$, $\hat B \equiv SBS^{-1}$, and $\hat C \equiv SCS^{-1}$. 
Note that $\hat A$ is diagonal 
with entries equal to the (real) eigenvalues. Let a dagger ($\dagger$) denote 
the Hermitian conjugate (complex conjugate $*$ plus transpose $T$). Since $\hat A^\dagger = \hat A$, we find 
\begin{eqnarray}\label{omegaeqn}
	(\omega - \omega^*) \hat v^\dagger \hat v & = & \hat v^\dagger (\omega \hat v) 
		- (\omega \hat v)^\dagger \hat v  \nonumber\\
		& = & \hat v^\dagger ( |k|\hat A - i\hat B - \hat C/|k|) \hat v 
		- \hat v^\dagger (|k| \hat A + i\hat B^T - \hat C^T/|k|) \hat v \nonumber\\
		& = & -i \hat v^\dagger (M + M^\dagger)\hat v
\end{eqnarray}
where $M \equiv \hat B - i\hat C/|k|$. The left--hand side includes the factor 
$(\omega - \omega^*) = 2i\Im\omega = -2i\Re(i\omega)$, so Eq.~(\ref{omegaeqn}) can be written as 
$2\Re(i\omega) = \hat v^\dagger (M + M^\dagger)\hat v/\hat v^\dagger \hat v$. 
It follows that $\Re(i\omega) \le \tau^{-1}$ 
where $2\tau^{-1}$ is the maximum over $|k|$ of the matrix norm of $(M + M^\dagger)$.
[The matrix norm is the maximum of the real number $\hat v^\dagger (M + M^\dagger)\hat v/(\hat v^\dagger \hat v)$.] 
From this argument we see that the growth 
rate for the mode $k_a$ is bounded; it can grow no faster than $e^{t/\tau}$ where $\tau$ is independent of $k_a$.

Consider a simple example in one spatial dimension with two pairs of canonically conjugate fields, 
$q_1$, $p_1$ and $q_2$, $p_2$. Let the Hamiltonian be given by
\begin{eqnarray}
	H & = & \int dx \biggl\{ \frac{1}{2}[ (p_1)^2 + (p_2)^2 + (q_1')^2 + (q_2')^2]  
		+ 2p_2 q_1'  + 2p_1q_2'  \nonumber\\
	& & \qquad\qquad + p_1(q_2 + q_1) + p_2(q_2 - q_1)  + q_1 q_2\biggr\} \ .
\end{eqnarray}
Hamilton's equations are
\bea
	\dot q_1 & = & p_1 + 2q_2' + q_2 + q_1 \ ,\\
	\dot q_2 & = & p_2 + 2q_1' + q_2 - q_1 \ ,\\
	\dot p_1 & = & q_1'' + 2p_2' + p_2 - p_1 - q_2 \ ,\\
	\dot p_2 & = & q_2'' + 2p_1' - p_1 - p_2  - q_1 \ .
\eea
In this example the PDE's are linear so the linearization step is trivial: $q_\mu\to \delta q_\mu$, 
$p_\mu \to \delta p_\mu$. 
Now insert the ansatz $\delta q_\mu = \bar q_\mu e^{i\omega t + ikx}/(i|k|)$, 
$\delta p_\mu = \bar p_\mu e^{i\omega t + ikx}$. 
This yields 
\bea
	\omega \bar q_1 & = & |k| (2n\bar q_2 + \bar p_1) - i(\bar q_1 + \bar q_2) \ ,\\
	\omega \bar q_2 & = & |k| (2n\bar q_1 + \bar p_2) - i(\bar q_2 - \bar q_1) \ ,\\
	\omega \bar p_1 & = & |k| (\bar q_1 + 2n\bar p_2) - i(\bar p_2 - \bar p_1) + \bar q_2/|k| \ ,\\
	\omega \bar p_2 & = & |k| (\bar q_2 + 2n\bar p_1) +i(\bar p_1 + \bar p_2) + \bar q_1/|k| \ ,
\eea
where $n \equiv k/|k|$ is the sign of the wave number $k$. 
Collecting the unknowns into a column vector $v = (\bar q_1,\bar q_2, \bar p_1, \bar p_2)^T$, we see that these 
equations become $\omega v = (|k| A -i B - C/|k|)v$ where the matrices are given by 
\be
	A =	\left( \begin{array}{cccc} 0 & 2n &  1 & 0 \\ 2n & 0 & 0 & 1 \\
 		  1 & 0 & 0 & 2n \\ 0 & 1  & 2n & 0 \end{array}\right)  \ ,\quad
	B = \left(\begin{array}{cccc} 1 & 1 & 0 & 0 \\ -1 & 1 & 0 & 0 \\
		 0 & 0 & -1 & 1 \\ 0 & 0 & -1 & -1 \end{array}\right) \ ,\quad
	C = \left(\begin{array}{cccc} 0 & 0 & 0 & 0 \\ 0 & 0 & 0 & 0 \\
		 0 & -1 & 0 & 0 \\ -1 & 0 & 0 & 0 \end{array} \right)  \ .
\ee
The principal symbol $A$ has real eigenvalues $\pm 1 $, $\pm 3$, and a complete set of eigenvectors. 
Therefore this system is strongly hyperbolic. 
The modes with wave number $k$ have frequencies $\omega  = \pm |k| + {\cal O}(1/|k|)$ and 
$\omega = \pm 3|k| + {\cal O}(1/|k|)$. In particular the 
imaginary parts of the $\omega$'s do not grow with increasing $|k|$.  

Now replace the terms $2p_2 q_1' + 2 p_1 q_2'$ in the Hamiltonian with $p_2 q_1' - p_1 q_2'$. The principal symbol becomes
\be
	A = \left( \begin{array}{cccc} 0 & -n &  1 & 0 \\ n & 0 & 0 & 1 \\  1 & 0 & 0 & n \\ 0 & 1  & -n & 0 \end{array}\right) 
\ee
while $B$ and $C$ are unchanged. The eigenvalues of $A$ vanish and there are only two independent eigenvectors. 
Therefore this system is weakly hyperbolic. The modes with wave number $k$ have frequencies 
$\omega = \pm i\sqrt{2|k|} + {\cal O}(1)$ and 
$\omega = \pm \sqrt{2|k|} + {\cal O}(1)$. The modes with frequency $\omega \approx -i\sqrt{2|k|}$ will grow in time at a rate 
that is unbounded as $|k|$ increases. 

The system described by this last example is not well posed. 
Indeed, consider two initial data sets that differ from one another by 
terms $q_\mu \sim \bar q_\mu e^{i k x}/|k|^2$, $p_\mu \sim \bar p_\mu e^{ikx}/|k|$ where 
$(\bar q_1,\bar q_2, \bar p_1, \bar p_2)^T$ is an eigenvector with eigenvalue 
$\omega = -i\sqrt{2|k|}+ {\cal O}(1)$.
In the limit as $|k|\to\infty$ these terms vanish and the 
two initial data sets coincide. However, if we evolve these  data sets the solutions will differ at finite time $t$ by terms 
$q_\mu \sim \bar q_\mu e^{\sqrt{2|k|}t + ikx}/|k|^2$, $p_\mu \sim \bar p_\mu e^{\sqrt{2|k|}t + ikx}/|k|$. 
These terms do not vanish in the limit $|k|\to\infty$. This system is 
ill posed because the solution at finite time does not depend continuously on the initial data. 

In some cases it may be possible to model a physical system with ill posed PDE's and to gain important physical 
insights through a formal analysis. 
Claudio's beautiful work on the (weakly hyperbolic) ADM equations is a perfect example! One can 
imagine that the initial data are analytic, in which case the Cauchy--Kowalewski theorem guarantees that a 
solution exists for a finite time. But most data, even smooth data, are not analytic. 
From a computational point of view, having an ill posed system is unacceptable. Numerical errors  
will always introduce modes with large wave numbers, with the size of 
$|k|$ limited only by the details of the numerical implementation. 
For example, with a finite difference algorithm the maximum $|k|$ is roughly $1/\Delta x$ where $\Delta x$ 
is the grid spacing. In practice it does not 
take long for the numerical solution to become dominated by this highest--wave number mode. As the grid resolution 
is increased ($\Delta x$ is decreased), the unwanted highest wave number mode grows even more quickly. 
For practical numerical studies, we need our system of PDE's to be well posed. 

The analysis outlined above leads to the following test for strong hyperbolicity. 
We begin by  constructing the principal symbol $A$ from the principal parts of 
Hamilton's equations. 
The principal parts of the $\dot q_\mu$ equations are the terms proportional to 
$p_\mu$ and   $\partial_a q_\mu$. In these terms we make the replacements $p_\mu \to 
\bar p_\mu$ and $\partial_a q_\mu \to n_a \bar q_\mu$. The principal parts of the $\dot p_\mu$ equations are 
the terms proportional to  $\partial_a p_\mu$ and  $\partial_a \partial_b q_\mu$. 
In these terms we make the replacements $\partial_a p_\mu \to n_a \bar p_\mu$ and 
$\partial_a \partial_b q_\mu \to n_a n_b \bar q_\mu$. The principal symbol $A$ is the matrix formed from
the coefficients of the $\bar q$'s and $\bar p$'s. The next step is to compute the eigenvalues and 
eigenvectors of $A$. If $A$ has real eigenvalues and a complete 
set of eigenvectors that depend smoothly on $n_a$, the system is strongly hyperbolic and the initial 
value problem is well posed. 

\section{ADM with dynamical lapse and shift}
The Einstein--Hilbert action is $S = \int d^4x \sqrt{-{\mathbf g}} {\mathbf R}$ where ${\mathbf g}$ 
is the determinant of the spacetime metric and
${\mathbf R}$ is the spacetime curvature scalar. Units are chosen such that Newton's constant 
equals $1/(16\pi)$. With the familiar 
splitting of the spacetime metric into the spatial metric $g_{ab}$, lapse function $\alpha$, 
and shift vector $\beta^a$, the action becomes 
\be\label{LagrangianAction}
	S[g,\alpha,\beta] = \int d^4x \, \alpha \sqrt{g} \left( R + K_{ab}K^{ab} - K^2 \right)  \ .
\ee
The extrinsic curvature is defined by 
\be\label{ExtrinsicCurvature}
	K_{ab} \equiv -\frac{1}{2\alpha} \left( \dot g_{ab} - 2 D_{(a}\beta_{b)} \right) \ ,
\ee
and $D_a$ denotes the spatial covariant derivative. The Hamiltonian can be derived in a straightforward fashion 
if one recognizes that the action does not contain time derivatives of the lapse and shift. Time derivatives 
of the spatial metric appear through the combination $K_{ab}$. Thus, we introduce the momentum 
\be\label{Momentum}
	P^{ab} \equiv \frac{\partial {\cal L}}{\partial {\dot g_{ab}}} = \sqrt{g} \left( K g^{ab} - K^{ab} \right) \ ,
\ee
where the Lagrangian density ${\cal L}$ is the integrand of the action. This definition can be inverted for $\dot g_{ab}$ as a 
function of $P^{ab}$ and used to define the Hamiltonian: $H \equiv \int d^3x \left( P^{ab} \dot g_{ab} - {\cal L} \right)$. 
This yields the ADM Hamiltonian 
\be 
	H = \int d^3x \left( \alpha {\cal H} + \beta^a {\cal M}_a \right) \ ,
\ee
where 
\bea
	{\cal H} & \equiv & \frac{1}{\sqrt{g}}\left( P^{ab}P_{ab} - P^2/2\right)  - \sqrt{g}R \ , \\
	{\cal M}_a & \equiv & -2 D_b P^b_a 
\eea
are the Hamiltonian and momentum constraints. 

In the analysis above the lapse and shift are treated as non dynamical fields. They appear in the 
Hamiltonian form of the action, 
\be
 	S[g, P, \alpha, \beta]  = \int_{t_i}^{t_f} dt \int d^3x \left\{ P^{ab} \dot g_{ab} 
		- \alpha {\cal H} - \beta^a {\cal M}_a  \right\} \ ,
\ee
as undetermined multipliers. Here $t_i$ and $t_f$ are the initial and final times. 
Extremization of $S$ with respect to $\alpha$ and $\beta^a$ yields the 
constraints ${\cal H} = 0$ and ${\cal M}_a = 0$. 

We can enlarge the ADM formulation to include the lapse and shift as dynamical 
variables.\cite{DeWitt:1967yk,HRT} Consider again the action (\ref{LagrangianAction}). 
In addition to the momentum $P^{ab}$ conjugate 
to the spatial metric, we also define conjugate variables for the lapse and shift:
\bea
	\pi & \equiv & \frac{\partial{\cal L}}{\partial \dot\alpha} = 0 \ ,\\
	\rho_a & \equiv & \frac{\partial{\cal L}}{\partial \dot \beta^a} = 0 \ .
\eea
This leads to primary constraints $\pi = 0$ and $\rho_a = 0$. The resulting Hamiltonian is not unique; it is 
only determined to within the addition of arbitrary multiples of the constraints: 
\be \label{Hamiltonian}
	H = \int d^3 x \left( \alpha {\cal H} + \beta^a {\cal M}_a 
		+ \Lambda \pi + \Omega^a \rho_a \right) \ .
\ee
The coefficients $\Lambda$ and $\Omega^a$ appear as undetermined multipliers in the action, which now reads
\begin{eqnarray}
	& & S[g,P,\alpha,\pi,\beta,\rho,\Lambda,\Omega] \nonumber\\ 
	& & \quad =  \int_{t_i}^{t_f} dt 
		\int d^3x \biggl\{ P^{ab}\dot g_{ab} + \pi \dot \alpha 
		+ \rho_a \dot \beta^a - \alpha{\cal H} - \beta^a {\cal M}_a 
		- \Lambda \pi - \Omega^a \rho_a \biggr\} \ .\qquad
\end{eqnarray}
The equations of motion, $\delta S =0$, are\footnote{Throughout this 
paper I ignore the issues that arise when space has boundaries.\cite{Regge:1974zd,Brown:actionenergy}}
\bea\label{enlargedADM}
	\dot g_{ab} & = & {\cal L}_\beta g_{ab} + \frac{\alpha}{\sqrt{g}} (2P_{ab} - Pg_{ab}) \ ,\slabel{ADMg}\\
	\dot P^{ab} & = & {\cal L}_\beta P^{ab} + \frac{\alpha}{\sqrt{g}} (\delta_c^a\delta_d^b 
	- g^{ab}g_{cd}/4)(P P^{cd} - 2P^{ce}P_e^d ) \nonumber\\
	& & \qquad - \alpha\sqrt{g} G^{ab} + \sqrt{g} (D^a D^b \alpha 
	- g^{ab} D_c D^c \alpha) \ ,\slabel{ADMP}\\
	\dot\alpha & = & \Lambda \ ,\slabel{ADMalpha}\\
	\dot \pi & = & -{\cal H} \ ,\slabel{ADMpi}\\
	\dot \beta^a & = & \Omega^a \ ,\slabel{ADMbeta}\\
	\dot \rho_a & = & -{\cal M}_a \ ,\slabel{ADMrho}\\
	\pi & = & 0 \ ,\slabel{ADMpiconstraint}\\
	\rho_a & = & 0 \ ,\slabel{ADMrhoconstraint}
\eea
where $G^{ab}$ denotes the spatial Einstein tensor and ${\cal L}_\beta$ is the Lie derivative with respect to $\beta^a$. 

The equations above must hold for each time $t_i \le t \le t_f$. 
They are equivalent to the Einstein equations supplemented with 
evolution equations for the lapse function and shift vector. In particular, observe that 
$\pi$ and $\rho_a$ must vanish for all time by Eqs.~(\ref{ADMpiconstraint},\ref{ADMrhoconstraint}). 
It follows that the time derivatives of $\pi$ and $\rho_a$  must vanish. In turn, 
Eqs.~(\ref{ADMpi},\ref{ADMrho}) imply that the usual Hamiltonian and momentum constraints are zero. 
Eqs.~(\ref{ADMg},\ref{ADMP}) are the familiar ADM evolution equations, 
and Eqs.~(\ref{ADMalpha},\ref{ADMbeta}) supply evolution equations for the lapse and shift. 

The evolution equations (\ref{enlargedADM}a--f) are Hamilton's equations derived from the Hamiltonian 
(\ref{Hamiltonian}). The time derivative of any function $F$ of the canonical variables is 
$\dot F = \{F,H\}$ where the fundamental Poisson brackets relations are defined by 
$\{g_{ab}(x),P^{cd}(x')\} = \delta_{(a}^{(c} \delta_{b)}^{d)}\delta^3(x,x')$, 
$\{\alpha(x),\pi(x')\} = \delta^3(x,x')$, and $\{\beta^a(x),\rho_b(x')\} = \delta^a_b \delta^3(x,x')$.
We can interpret Hamilton's equations as an initial value problem by following Dirac's 
reasoning for constrained Hamiltonian systems.\cite{HenneauxTeitelboim}
The initial data are chosen such that the primary constraints $\pi$ and $\rho_a$ vanish 
at the initial time. According to Eqs.~(\ref{ADMpi},\ref{ADMrho}), these constraints  will remain 
zero as long as ${\cal H}$ and ${\cal M}_a$ are constrained to   
vanish as well. Thus we impose ${\cal H} = 0$ and ${\cal M}_a=0$ as secondary constraints. 
The complete set of constraints, $\pi=0$, $\rho_a=0$, ${\cal H}=0$, and ${\cal M}_a=0$ is first class. 

\section{Hyperbolicity of ADM with dynamical lapse and shift}
Hamilton's equations (\ref{enlargedADM}a--f) are equivalent to the ADM equations plus evolution equations 
$\dot\alpha = \Lambda$, $\dot\beta^a = \Omega^a$ for the lapse and shift. Let us analyze the level of hyperbolicity
of these PDE's. 
The principal parts of the $\dot q$ equations are the terms that are proportional to $p$'s or first spatial 
derivatives of $q$'s. The principal parts of the $\dot p$ equations are the terms that are proportional 
to first spatial derivatives of $p$'s or second spatial derivatives of $q$'s. Thus, we find 
\bea\label{ADMprincipal}
	\dnot g_{ab} & \cong & 2g_{c(a}\partial_{b)}\beta^c + \frac{\alpha}{\sqrt{g}}(2P_{ab} - Pg_{ab}) \ ,\\
	\dnot P^{ab} & \cong & \frac{\alpha \sqrt{g}}{2} g^{ac}g^{bd} g^{ef} (\partial_e\partial_f g_{cd} 
  		- 2 \partial_e\partial_{(c} g_{d)f} + \partial_c\partial_d g_{ef}) \nonumber\\
		& & + \frac{\alpha\sqrt{g}}{2} g^{ab} g^{cd} g^{ef} (\partial_c\partial_e g_{df} 
		- \partial_c\partial_d g_{ef}) \nonumber\\
		& & + \sqrt{g} (g^{ac}g^{bd} - g^{ab} g^{cd})\partial_c\partial_d \alpha \ ,\\
	\dnot \alpha & \cong & -\beta^a \partial_a\alpha \ ,\\
	\dnot \pi & \cong & \sqrt{g} g^{ab}g^{cd}(\partial_a\partial_c g_{bd} - \partial_a\partial_b g_{cd}) 
		- \beta^a \partial_a \pi \ ,\\
	\dnot \beta^a & \cong & -\beta^b\partial_b \beta^a \ ,\\
	\dnot \rho_a & \cong & 2 g_{ac}\partial_b P^{bc} - \beta^b \partial_b \rho_a \ ,
\eea
where the symbol $\cong$ is used to denote equality 
up to lower order (non principal) terms.  These equations have been expressed in terms of the operator 
$\dnot \equiv \partial_t - \beta^a \partial_a$ so that the characteristic speeds (the eigenvalues of the 
principal symbol) are defined with respect to observers who are at rest in the spacelike slices. 

We now construct the eigenvalue problem $\mu v = A v$ for the principal symbol $A$. 
The principal symbol is found from 
the coefficients on the right--hand sides of Eqs.~(\ref{ADMprincipal}). These coefficients are divided by 
a factor of $\alpha$ so that the characteristic speeds will be expressed in terms of proper time rather 
than coordinate time. The result is  
\bea\label{ADMevproblem}
	\mu \bar g_{ab} & = & \frac{2}{\alpha} n_{(a}\bar\beta_{b)} + \frac{1}{\sqrt{g}} [ 2\bar P_{ab} 
		- g_{ab} (\bar P_{nn} + \bar P_{AB}\delta^{AB}) ] \ ,\\
	\mu \bar P_{ab} & = & \frac{\sqrt{g}}{2} [ \bar g_{ab} - 2n_{(a}\bar g_{b)n} 
		+ n_a n_b (\bar g_{nn} + \bar g_{AB}\delta^{AB} ) - g_{ab} \bar g_{AB}\delta^{AB} ] 
		\nonumber\\
		& & - \frac{\sqrt{g}}{\alpha}(g_{ab} - n_a n_b) \bar\alpha \ ,\\
	\mu \bar\alpha & = & -(\beta\cdot n/\alpha)\bar\alpha \ ,\\
	\mu\bar\pi  & = & -\frac{\sqrt{g}}{\alpha} \bar g_{AB}\delta^{AB} - (\beta\cdot n/\alpha)\bar\pi \ ,\\
	\mu\bar\beta_a & = & -(\beta\cdot n/\alpha)\bar\beta_a \ ,\\
	\mu\bar\rho_a & = & \frac{2}{\alpha} \bar P_{na} - (\beta\cdot n/\alpha)\bar\rho_a \ .
\eea
where $\mu$ is the eigenvalue and $v = (\bar g_{ab},\bar P_{ab},\bar\alpha,\bar\pi,\bar\beta_a,\bar\rho_a)^T$ 
is the eigenvector. In these equations $n^a$ is normalized with respect to the spatial metric, 
$n^a g_{ab} n^b = 1$, and 
a subscript $n$ denotes contraction with $n^a$. We have also introduced an orthonormal 
diad $e^a_A$ with $A = 1,2$  in the subspace orthogonal to $n_a$. That is, $n_a e^a_A = 0$ and 
$e^a_A g_{ab} e^b_B = \delta_{AB}$. A subscript $A$ on a tensor (such as the metric $g_{ab}$ or 
momentum $P_{ab}$) denotes contraction with $e^a_A$. 

The eigenvalue problem (\ref{ADMevproblem}) 
splits into scalar, vector and trace--free tensor blocks with respect to rotations about the 
normal direction $n_a$. The scalar block is
\bea
	\mu \bar g_{nn} & = & \frac{2}{\alpha}\bar\beta_n + \frac{1}{\sqrt{g}} (\bar P_{nn} 
		- \bar P_{AB}\delta^{AB}) \ ,\\
	\mu\bar g_{AB}\delta^{AB} & = & -\frac{2}{\sqrt{g}} \bar P_{nn} \ ,\\
	\mu\bar P_{nn} & = & 0 \ ,\\
	\mu\bar P_{AB}\delta^{AB} & = & -\frac{1}{2} \sqrt{g} \bar g_{AB}\delta^{AB} 
		- \frac{2}{\alpha}\sqrt{g} \bar \alpha \ ,\\
	\mu\bar\alpha & = & -(\beta\cdot n/\alpha)\bar\alpha \ ,\\
	\mu\bar\pi & = & -\frac{\sqrt{g}}{\alpha} \bar g_{AB}\delta^{AB} - (\beta\cdot n/\alpha)\bar\pi \ ,\\
	\mu\bar\beta_n & = & -(\beta\cdot n/\alpha)\bar\beta_n \ ,\\
	\mu\bar\rho_n & = & \frac{2}{\alpha} \bar P_{nn} - (\beta\cdot n/\alpha)\bar\rho_n \ .
\eea
This block has eigenvalues $0$ and $-(\beta\cdot n/\alpha)$, each with multiplicity 4. There 
is only one eigenvector with eigenvalue $0$, so the eigenvectors are not complete. 
The vector block is
\bea
	\mu \bar g_{nA} & = & \frac{1}{\alpha} \bar \beta_A + \frac{2}{\sqrt{g}} \bar P_{nA} \ ,\\
	\mu \bar P_{nA} & = & 0 \ ,\\
	\mu \bar \beta_A & = & -(\beta\cdot n/\alpha)\bar\beta_A \ ,\\
	\mu \bar\rho_A & = & \frac{2}{\alpha} \bar P_{nA} - (\beta\cdot n/\alpha)\bar\rho_A \ .
\eea
It has eigenvalues $0$ and $-(\beta\cdot n/\alpha)$, each with multiplicity 2. There is only one 
eigenvector with eigenvalue $0$, so the eigenvectors are not complete. 
Finally, the trace--free tensor block is
\bea
	\mu \bar g_{AB}^{tf} & = & \frac{2}{\sqrt{g}} \bar P_{AB}^{tf} \ ,\\
	\mu \bar P_{AB}^{tf} & = & \frac{\sqrt{g}}{2} \bar g_{AB}^{tf} \ .
\eea
This block has eigenvalues $\pm 1$ and a complete set of eigenvectors. 
Because the eigenvectors for the scalar and vector blocks are not complete, the system 
(\ref{enlargedADM}) is weakly hyperbolic. 

\section{Extending the ADM formulation}
In the previous section we modified the ADM Hamiltonian formulation of general relativity so that 
the lapse function $\alpha$ and shift vector $\beta^a$ are treated as dynamical variables. 
Their canonical conjugates are denoted $\pi$ and $\rho_a$, respectively. The undetermined multipliers 
for the constraints $\pi=0$, $\rho_a=0$ are $\Lambda$ and $\Omega^a$. These multipliers are 
freely specifiable functions of space and time. They determine the slicing and spatial coordinate 
conditions through the equations of motion $\dot\alpha = \Lambda$ and $\dot\beta^a = \Omega^a$. 

Here is the key observation. The histories that extremize the action are unchanged if we replace 
the multipliers by 
$\Lambda \to \Lambda + \hat\Lambda$ and $\Omega^a \to \Omega^a + \hat\Omega^a$, where 
$\hat\Lambda$ and $\hat\Omega^a$ are quasilinear functions of the canonical variables. 
By quasilinear, I mean that 
the principal parts of $\hat\Lambda$ and $\hat\Omega^a$ are linear in the momenta ($P^{ab}$, 
$\pi$ and $\rho_a$) and first spatial derivatives of the coordinates ($\partial_c g_{ab}$, 
$\partial_c\alpha$ and $\partial_c\beta^a$) with 
coefficients that depend on the coordinates. With these replacements the action becomes 
\begin{eqnarray}\label{xaction}
	S[g,P,\alpha,\pi,\beta,\rho,\Lambda,\Omega]  
  	& = &  \int_{t_i}^{t_f} dt \int d^3x \biggl\{ P^{ab}\dot g_{ab} 
		+ \pi \dot \alpha 
		+ \rho_a \dot \beta^a - \alpha{\cal H} - \beta^a {\cal M}_a \nonumber\\
	& & \qquad\qquad\qquad - (\Lambda + \hat\Lambda) \pi 
		- (\Omega^a + \hat\Omega^a) \rho_a \biggr\} \ ,
\end{eqnarray}
and the Hamiltonian is 
\be\label{xHamiltonian}
	H = \int d^3 x \left( \alpha {\cal H} + \beta^a {\cal M}_a +( \Lambda + \hat\Lambda) \pi 
		+( \Omega^a + \hat\Omega^a)  \rho_a \right) \ .
\ee
The solutions to the equations of motion are unaltered because the extra terms are proportional 
to the constraints $\pi = 0$, $\rho_a = 0$. In the Hamiltonian formulation we can dispense with the 
original multipliers $\Lambda$ and $\Omega^a$; that is, these quantities can be absorbed into 
the functions $\hat\Lambda$ and $\hat\Omega^a$. 

With  $\hat\Lambda$ and $\hat\Omega^a$ restricted to be  
quasilinear in the momenta and first spatial derivatives of the coordinates, the equations of motion become 
\bea
	\dot g_{ab} & = & {\cal L}_\beta g_{ab} + \frac{\alpha}{\sqrt{g}} (2P_{ab} - Pg_{ab}) + 
		\frac{\partial \hat\Lambda}{\partial P^{ab}} \pi 
		+ \frac{\partial\hat\Omega^c}{\partial P^{ab}} \rho_c \ ,\slabel{xADMg}\\
	\dot P^{ab} & = & {\cal L}_\beta P^{ab} + \frac{\alpha}{\sqrt{g}} (\delta_c^a\delta_d^b - g^{ab}g_{cd}/4)
		(P P^{cd} - 2P^{ce}P_e^d ) \nonumber\\
		& & - \alpha\sqrt{g} G^{ab} + \sqrt{g} (D^a D^b \alpha - g^{ab} D_c D^c \alpha) \nonumber\\
		& & - \frac{\partial\hat\Lambda}{\partial g_{ab}} \pi 
		- \frac{\partial\hat\Omega^c}{\partial g_{ab}} \rho_c 
		+ \partial_d \left( \frac{\partial\hat\Lambda}{\partial(\partial_d g_{ab})} \pi\right) 
		+ \partial_d \left( \frac{\partial\hat\Omega^c}{\partial(\partial_d g_{ab})} \rho_c \right) 
		\ , \slabel{xADMP}\\
	\dot\alpha & = & \Lambda + \hat\Lambda + \frac{\partial\hat\Lambda}{\partial\pi} \pi 
		+ \frac{\partial\hat\Omega^c}{\partial\pi}\rho_c  \ , \slabel{xADMalpha}\\
	\dot \pi & = & -{\cal H} - \frac{\partial\hat\Lambda}{\partial\alpha}\pi 
		- \frac{\partial\hat\Omega^c}{\partial\alpha} \rho_c 
		+ \partial_d \left( \frac{\partial\hat\Lambda}{\partial(\partial_d\alpha)}\pi\right) 
		+ \partial_d \left( \frac{\partial\hat\Omega^c}{\partial(\partial_d\alpha)}\rho_c\right) 
		\ , \slabel{xADMpi}\\
	\dot \beta^a & = & \Omega^a + \hat\Omega^a + \frac{\partial\hat\Lambda}{\partial\rho_a}\pi 
		+ \frac{\partial\hat\Omega^c}{\partial\rho_a}\rho_c \ , \slabel{xADMbeta}\\
	\dot \rho_a & = & -{\cal M}_a - \frac{\partial\hat\Lambda}{\partial\beta^a}\pi
		- \frac{\partial\hat\Omega^c}{\partial\beta^a}\rho_c 
		+ \partial_d \left( \frac{\partial\hat\Lambda}{\partial(\partial_d\beta^a)}\pi\right) 
		+ \partial_d \left( \frac{\partial\hat\Omega^c}{\partial(\partial_d\beta^a)}\rho_c\right)
		\ , \slabel{xADMrho}\\
	\pi & = & 0 \ , \\
	\rho_a & = & 0 \ , 
\eea
Equations (\ref{xADMg},\ref{xADMP}) are the usual ADM equations apart from terms 
proportional to the constraints, $\pi = 0$ and $\rho_a = 0$. The equations that govern the slicing and 
spatial coordinates are generalized by the presence of the functions $\hat\Lambda$ and $\hat\Omega^a$. 
Apart from terms that vanish with the constraints $\pi = 0$, 
$\rho_a = 0$, the evolution equation for the lapse becomes $\dot\alpha = \Lambda + \hat\Lambda$ and 
the evolution equation for the shift becomes $\dot\beta^a = \Omega^a + \hat\Omega^a$. The equations 
for $\dot\pi$ and $\dot\rho_a$ are modified, but once again we see that the complete set of 
constraints $\pi=0$, $\rho_a=0$, ${\cal H}=0$, and ${\cal M}_a =0$ is first class. 

In principle we can choose $\hat\Lambda$ and $\hat\Omega^a$ to be any set of quasilinear functions of the 
canonical variables. In practice we might want $\hat\Lambda$ and $\hat\Omega^a$ to 
satisfy certain transformation 
properties. For example we can restrict $\hat\Lambda$ to be a scalar and $\hat\Omega^a$ to be a 
contravariant vector under spatial diffeomorphisms. This allows us to maintain a 
geometrical interpretation of the equations of motion. In particular this allows us to prepare and 
evolve identical geometrical data using different spatial coordinate systems. 

Another property that can be imposed on the formalism 
is reparametrization invariance.\cite{HenneauxTeitelboim} This is 
invariance under a change of coordinate labels $t$ for the constant 
time slices. Consider the infinitesimal transformation $t \to t - \varepsilon(t)$. In the usual ADM system, the 
variables $g_{ab}$ and $P^{ab}$ transform as scalars under time reparamerization: 
$\delta g_{ab} = \varepsilon \dot g_{ab}$ and 
$\delta P^{ab} = \varepsilon \dot P^{ab}$. The time derivative 
of the metric, $\dot g_{ab}$, transforms as a covariant vector. In one dimension a covariant vector 
transforms in the same way as a scalar density of weight $+1$. It follows that the term $P^{ab} \dot g_{ab}$ 
that appears in the action is a weight $+1$ scalar density. For reparametrization invariance to hold, 
the integrand of the action should transform as a weight $+1$ scalar 
density since it is integrated over $t$. In particular the lapse function $\alpha$ and 
shift vector $\beta^a$, which multiply the scalars ${\cal H}$ and ${\cal M}_a$, 
must transform as scalar densities of weight $+1$. 

Observe that 
the time derivatives $\dot\alpha$ and $\dot\beta^a$ are constructed from coordinate derivatives of scalar 
densities and, as a consequence, these quantities do not transform as tensors or tensor densities. We need to 
replace the coordinate derivatives (dots) with covariant derivatives. We can do this by 
choosing a background metric for the time direction. This should be viewed as part of the gauge fixing 
process. 
Now, the physical metric for the time manifold is $\alpha^2$, so let $\tilde\alpha^2$ denote the background metric. 
The covariant derivative built from $\tilde\alpha^2$, acting on the densities $\alpha$ and $\beta^a$, 
is defined by 
\bea
	\mathring\alpha & \equiv & \dot\alpha - (\dot{\tilde\alpha}/\tilde\alpha) \alpha \ ,\\
	\mathring\beta^a & \equiv & \dot\beta^a - (\dot{\tilde\alpha}/\tilde\alpha)\beta^a \ .
\eea
The extra terms needed for reparametrization invariance can be built into the action or Hamiltonian 
by including a term $(\dot{\tilde\alpha}/\tilde\alpha) \alpha$ in the function 
$\hat\Lambda$ and a term $(\dot{\tilde\alpha}/\tilde\alpha)\beta^a$ in the function $\hat\Omega^a$. 

With the appropriate terms included in $\hat\Lambda$ and $\hat\Omega^a$, the time derivatives of the 
lapse and shift appear in the action only in the combinations $\pi\mathring\alpha$ and $\rho_a\mathring\beta^a$. 
Since $\mathring\alpha$ and $\mathring\beta^a$ are covariant vector densities of weight $+1$, we see 
that $\pi$ and $\rho_a$ must transform as contravariant vectors with no density weight. In one dimension, 
contravariant vectors transform in the same way as a scalar density of weight $-1$. We will 
consider $\pi$ and $\rho_a$ to be scalar densities of weight $-1$ under time reparametrization. It follows 
that, apart from the terms $(\dot{\tilde\alpha}/\tilde\alpha) \alpha$ and 
$(\dot{\tilde\alpha}/\tilde\alpha)\beta^a$, the multipliers $\Lambda + \hat\Lambda$ and 
$\Omega^a + \hat\Omega^a$ should transform as scalar densities of weight $+2$. 

We have now established the rules for adding terms to the functions $\hat\Lambda$ and $\hat\Omega^a$ such 
that the resulting formulation is invariant under time reparametrizations: these terms must 
be weight $+2$ densities built 
from the scalars $g_{ab}$, $P^{ab}$, the weight $+1$ densities 
$\alpha$, $\beta^a$, and the weight $-1$ densities $\pi$, $\rho_a$. We can also insist that $\hat\Lambda$ 
and $\hat\Omega^a$ should be, respectively, a scalar and a contravariant vector under spatial diffeomorphisms. 
With these properties in mind, a fairly general form for $\hat\Lambda$ is 
\be
	\hat\Lambda = (\dot{\tilde\alpha}/\tilde\alpha)\alpha + \beta^a D_a\alpha 
		- C_1 \frac{\alpha^2}{\sqrt{g}} P 
		+ C_4 \frac{\alpha^3}{\sqrt{g}} \pi \ .
\ee
The first term in required for reparametrization invariance. The second term will allow us to 
combine shift vector terms into a Lie derivative ${\cal L}_\beta$ acting on $\alpha$. 
The terms multiplied by constants $C_1$ and 
$C_4$ are principal terms that will affect the hyperbolicity of the resulting equations. There 
are other principal terms that one can add, such as $\alpha^2\beta^a\rho_a/\sqrt{g}$, but the form above will be 
general enough for present purposes. There are also lower order terms that one can add to 
$\hat\Lambda$. 

For $\hat\Omega^a$ we must construct a spatial vector that [apart from the term 
$(\dot{\tilde\alpha}/\tilde\alpha)\beta^a$] transforms as a weight $+2$ density under 
time reparametrizations. There are several obvious ways to construct a spatial vector from the 
canonical variables at hand. There are some possibilities that are not so obvious. Recall that the 
difference of two connections is a tensor. Thus, the combination $\Gamma^a_{bc} - \tilde\Gamma^a_{bc}$ 
is a spatial tensor if $\Gamma^a_{bc}$ are the Christoffel symbols built from the physical 
metric $g_{ab}$ and $\tilde\Gamma^a_{bc}$ 
are the Christoffel symbols built from a background metric $\tilde g_{ab}$.  
In setting up a numerical calculation, for example,  
on a logically Cartesian grid, we can take $\tilde g_{ab}$ to be the flat metric in Cartesian coordinates.  
Again, we view the introduction of the background structure $\tilde g_{ab}$ as part of the gauge 
fixing process. 

The general form for $\hat\Omega^a$ that we will consider is 
\begin{eqnarray}
	\hat\Omega^a  & = & (\dot{\tilde\alpha}/\tilde\alpha)\beta^a 
		+ \beta^b {\tilde D}_b\beta^a 
		+ C_2 \alpha^2 (\Gamma^a_{bc} - \tilde\Gamma^a_{bc}) g^{bc} \nonumber\\
		& & 
		+ C_3 \alpha^2 (\Gamma^c_{cb}  - \tilde\Gamma^c_{cb})g^{ab} 
		- C_5 \alpha D^a\alpha - C_6 \frac{\alpha^3}{\sqrt{g}} \rho^a \ .
\end{eqnarray}
where ${\tilde D}_a$ is the covariant derivative compatible with $\tilde g_{ab}$. 
The first term is required for time reparametrization invariance. The second 
term will allow us to combine time derivatives and shift vector terms into the operator 
$\dnot$ in the principle parts of the equations for $\beta^a$ and $\rho_a$.  
The remaining terms will modify the principal parts of the equations of motion and can affect the 
hyperbolicity of the system. There are other principal terms that we could add to
$\hat\Omega^a$, such as 
$\alpha^2 \pi\beta^a/\sqrt{g}$ or $\alpha P^{ab}\beta_b/\sqrt{g}$. We can also add 
lower order terms. 

With these expressions for $\hat\Lambda$ and $\hat\Omega^a$, we find the following equations of motion 
by varying the action (\ref{xaction}): 
\bea\label{cADM}
	\mathring g_{ab} & = & {\cal L}_\beta g_{ab} + \frac{\alpha}{\sqrt{g}} (2P_{ab} - Pg_{ab}) 
		- C_1 \frac{\alpha^2}{\sqrt{g}} \pi g_{ab}  \ ,\label{cADMg}\\
	\mathring P^{ab} & = & {\cal L}_\beta P^{ab} + \frac{\alpha}{\sqrt{g}} (\delta_c^a\delta_d^b - g^{ab}g_{cd}/4)
		(P P^{cd} - 2P^{ce}P_e^d )
		- \alpha\sqrt{g} G^{ab} + \nonumber\\ & & 
		\sqrt{g} (D^a D^b \alpha - g^{ab} D_c D^c \alpha) 
		+ C_1 \frac{\alpha^2}{\sqrt{g}} (P^{ab} - Pg^{ab}/2) \pi 
		+ C_4\frac{\alpha^3}{2\sqrt{g}} \pi^2 g^{ab} \nonumber\\ & & 
		+ C_2 D^{(a} (\rho^{b)}\alpha^2) - C_5 \alpha \rho^{(a} D^{b)}\alpha  
		+ \frac{1}{2}(C_3 - C_2) D_c(\alpha^2\rho^c) g^{ab}  \nonumber\\ & & 
		 + C_2 \alpha^2 \rho_e (\Gamma^e_{cd} - \tilde\Gamma^e_{cd})g^{ac} g^{bd} 
		+ C_3 \alpha^2 (\Gamma^d_{cd} - \tilde\Gamma^d_{cd}) \rho^{(a} g^{b)c}  
		\nonumber\\ & & 
		- C_6 \frac{\alpha^3}{\sqrt{g}} (\rho^a\rho^b + \rho_c\rho^c g^{ab}/2) 
		\ , \label{cADMP}\\
	\mathring\alpha & = & {\cal L}_\beta \alpha + \Lambda 
			- C_1\frac{\alpha^2}{\sqrt{g}} P + 2 C_4 \frac{\alpha^3}{\sqrt{g}} \pi 
		 \ , \label{cADMalpha}\\
	\mathring \pi & = & {\cal L}_\beta \pi -{\cal H} + 2C_1\frac{\alpha}{\sqrt{g}} P\pi 
		- 3C_4 \frac{\alpha^2}{\sqrt{g}} \pi^2 
		- 2C_2\alpha (\Gamma^a_{bc} - \tilde\Gamma^a_{bc}) g^{bc}\rho_a  \nonumber\\
		& & - 2C_3\alpha (\Gamma^b_{ab} - \tilde\Gamma^b_{ab})\rho^a 
		- C_5\alpha D_a\rho^a + 3 C_6 \frac{\alpha^2}{\sqrt{g}} \rho_a\rho^a
		\ , \label{cADMpi}\\
	\mathring \beta^a & = &  \beta^b {\tilde D}_b \beta^a + \Omega^a 
		+ C_2\alpha^2 (\Gamma^a_{bc} - \tilde\Gamma^a_{bc})g^{bc} 
		+ C_3\alpha^2 (\Gamma^c_{bc} - \tilde\Gamma^c_{bc})g^{ab} \nonumber\\ & & 
		- C_5\alpha D^a\alpha 
		- 2C_6 \frac{\alpha^3}{\sqrt{g}} \rho^a \ , \label{cADMbeta}\\
	\mathring \rho_a & = &  {\tilde D}_b(\beta^b\rho_a) - \rho_b {\tilde D}_a\beta^b 
		-{\cal M}_a - \pi D_a\alpha
		\ , \label{cADMrho}\\
	\pi & = & 0 \ , \\
	\rho_a & = & 0 \ .
\eea
These equations are generally covariant under spatial diffeomorphisms and time 
reparametrizations. 
Equations (\ref{cADM}a--f) are generated through the Poisson brackets by the Hamiltonian (\ref{xHamiltonian}). 
Note that for $g_{ab}$ and $P^{ab}$, which are scalars under time reparametrization, the 
covariant time derivative (circle) is equivalent to a coordinate time derivative (dot). 

\section{Hyperbolicity of the Extended ADM formulation}
The principal parts of the extended ADM evolution equations (\ref{cADM}a--f) are:
\bea
	\dnot g_{ab} & \cong & 2g_{c(a}\partial_{b)}\beta^c + \frac{\alpha}{\sqrt{g}}(2P_{ab} - Pg_{ab}) 
		- C_1\frac{\alpha^2}{\sqrt{g}} \pi g_{ab} \ ,\\
	\dnot P^{ab} & \cong & \frac{\alpha \sqrt{g}}{2} g^{ac}g^{bd} g^{ef} (\partial_e\partial_f g_{cd} 
  		- 2 \partial_e\partial_{(c} g_{d)f} + \partial_c\partial_d g_{ef}) \nonumber\\
		& & + \frac{\alpha\sqrt{g}}{2} g^{ab} g^{cd} g^{ef} (\partial_c\partial_e g_{df} 
		- \partial_c\partial_d g_{ef})
		+ \sqrt{g} (g^{ac}g^{bd} - g^{ab} g^{cd})\partial_c\partial_d \alpha  \nonumber\\
		& & +\alpha^2 \left[ C_2 g^{c(a}g^{b)d} 
		+ (C_3 - C_2) g^{ab} g^{cd}/2 \right] \partial_c\rho_d   \ ,\\
	\dnot \alpha & \cong & -C_1\frac{\alpha^2}{\sqrt{g}} P + 2C_4 \frac{\alpha^3}{\sqrt{g}} \pi  \ ,\\
	\dnot \pi & \cong & \sqrt{g} g^{ab}g^{cd}(\partial_a\partial_c g_{bd} - \partial_a\partial_b g_{cd}) 
		-  C_5 \alpha g^{ab}\partial_a\rho_b   \ ,\\
	\dnot \beta^a & \cong & \alpha^2 \left[ C_2 g^{ac} g^{bd}  
		+ (C_3 - C_2) g^{ab} g^{cd}/2\right]  \partial_b g_{cd} \nonumber\\ & & 
		- C_5 \alpha g^{ab} \partial_b\alpha - 2C_6 \frac{\alpha^3}{\sqrt{g}} \rho^a  \ ,\\
	\dnot \rho_a & \cong & 2 g_{ac}\partial_b P^{bc}   \ ,
\eea
From here it is easy to construct the eigenvalue problem for the principal symbol. The symbol decomposes 
into scalar, vector, and trace--free tensor blocks under rotations orthogonal to the normal vector 
$n_a \equiv k_a/|k|$. For the scalar sector, we find 
\bea
	\mu \bar g_{nn} & = & \frac{2}{\alpha}\bar\beta_n + \frac{1}{\sqrt{g}} (\bar P_{nn} - \bar P_{AB}\delta^{AB}) 
		- C_1 \frac{\alpha}{\sqrt{g}} \bar\pi  \ ,\\
	\mu\bar g_{AB}\delta^{AB} & = & -\frac{2}{\sqrt{g}} \bar P_{nn} 
		- 2C_1 \frac{\alpha}{\sqrt{g}} \bar\pi \ ,\\
	\mu\bar P_{nn} & = &  \frac{\alpha}{2}(C_3 + C_2) \bar\rho_n \ ,\\
	\mu\bar P_{AB}\delta^{AB} & = & -\frac{1}{2} \sqrt{g} \bar g_{AB}\delta^{AB} 
		- \frac{2}{\alpha}\sqrt{g} \bar \alpha + \alpha (C_3 - C_2) \bar\rho_n \ ,\\
	\mu\bar\alpha & = & -C_1 \frac{\alpha}{\sqrt{g}} (\bar P_{nn} + \bar P_{AB}\delta^{AB}) 
		+ 2C_4 \frac{\alpha^2}{\sqrt{g}} \bar\pi  \ ,\\
	\mu\bar\pi & = & -\frac{\sqrt{g}}{\alpha} \bar g_{AB}\delta^{AB} - C_5 \bar\rho_n  \ ,\\
	\mu\bar\beta_n & = & \frac{\alpha}{2}(C_3 + C_2) \bar g_{nn} 
		+ \frac{\alpha}{2}(C_3 - C_2) \bar g_{AB}\delta^{AB} 
		-  C_5 \bar\alpha - 2C_6 \frac{\alpha^2}{\sqrt{g}} \bar\rho_n  \ ,\quad\\
	\mu\bar\rho_n & = & \frac{2}{\alpha} \bar P_{nn}  \ .
\eea
Again, the subscripts $n$ and $A$ denote contraction with $n^a$ and $e^a_A$, respectively. 
The vector block is 
\bea
	\mu \bar g_{nA} & = & \frac{1}{\alpha}\bar \beta_A + \frac{2}{\sqrt{g}} \bar P_{nA} \ ,\\
	\mu \bar P_{nA} & = & \frac{\alpha}{2} C_2 \bar\rho_A \ ,\\
	\mu \bar \beta_A & = & C_2\alpha \bar g_{nA} - 2C_6 \frac{\alpha^2}{\sqrt{g}} \bar\rho_A \ ,\\
	\mu \bar\rho_A & = & \frac{2}{\alpha} \bar P_{nA} \ ,
\eea
and the trace--free tensor block is unmodified from before: 
\bea
	\mu \bar g_{AB}^{tf} & = & \frac{2}{\sqrt{g}} \bar P_{AB}^{tf} \ ,\\
	\mu \bar P_{AB}^{tf} & = & \frac{\sqrt{g}}{2} \bar g_{AB}^{tf} \ .
\eea
The eigenvalues for the scalar block are $\pm \sqrt{2C_1}$ and $\pm \sqrt{C_2 + C_3}$, 
each with multiplicity two. The eigenvalues for the vector block are $\pm\sqrt{C_2}$, each with 
multiplicity two. For the tensor block the eigenvalues are $\pm 1$ and the eigenvectors 
are complete. 

The eigenvalues are the characteristic speeds with respect to observers at 
rest in the spacelike hypersurfaces. 
Let us choose the $C$'s such that the characteristics are $\pm 1$, that is, along the physical 
light cone. Then we must have $C_1 = 1/2$, $C_2 = 1$, and $C_3 = 0$. With this choice a careful analysis of the 
scalar block shows that the eigenvectors are complete only if $C_4 = 1/8$, $C_5 = 1$, and 
$C_6 = 1/2$. For these values of the constants the eigenvectors for the vector block are complete as well. 
It can be shown that with these values for the $C$'s the principal parts of the equations of motion are equivalent to the 
generalized harmonic formulation of relativity.\cite{Friedrich:1985,Pretorius:2004jg,Lindblom:2005qh} 
This will be discussed elsewhere.\cite{BrownPrep}

The Hamiltonian system (\ref{cADM}) is strongly hyperbolic with the choice of $C$'s above. 
For this formulation the gauge conditions are  
\bea 
	\mathring \alpha & = & \beta^a D_a\alpha + \Lambda 
		- \frac{\alpha^2}{2} \frac{P}{\sqrt{g}} + \frac{\alpha^3}{4} \frac{\pi}{\sqrt{g}} \ ,\\
	\mathring \beta^a & = & \beta^b {\tilde D}_b\beta^a + \Omega^a 
		+ \alpha^2 (\Gamma^a_{bc} - \tilde\Gamma^a_{bc}) g^{bc} 
		- \alpha D^a \alpha - \frac{\alpha^3}{\sqrt{g}} \rho^a 
\eea 
If we choose  $\Lambda$ and $\Omega^a$ to vanish,
these gauge equations become 
\bea\label{GHgauge}
	\dot\alpha - \beta^a\partial_a\alpha & = & \alpha (\dot{\tilde\alpha}/\tilde\alpha)  - \alpha^2 K \ ,\\
	\dot\beta^a - \beta^b\partial_b\beta^a & = & \beta^a (\dot{\tilde\alpha}/\tilde\alpha)  +  \alpha^2 
		( \Gamma^a_{bc} - \tilde\Gamma^a_{bc})g^{bc} - \alpha g^{ab} \partial_b \alpha  \ ,
\eea
to within terms that vanish with the constraints $\pi = 0$, $\rho_a=0$. Here, 
$K \equiv P/(2\sqrt{g})$ is the trace of the extrinsic curvature. These gauge conditions agree, to within 
lower order (non--principal) terms, with the gauge conditions for the generalized harmonic formulation 
of gravity.\cite{Lindblom:2005qh}

It is not difficult to find choices for the constants (the $C$'s) that make the system (\ref{cADM}) strongly 
hyperbolic. The trick is to find a strongly hyperbolic system with desirable coordinate and slicing 
conditions. Conditions similar to (\ref{GHgauge}), supplemented with ``gauge driver" equations, 
have been applied to the binary black hole problem with mixed 
results.\cite{Pretorius:2004jg,Lindblom:2005qh,Lindblom:2007xw} 
With the BSSN evolution system, the gauge conditions that work well for the binary black hole problem 
are ``1+log slicing",  
\be\label{onepluslog}
	\dot\alpha - \beta^a\partial_a\alpha = - 2\alpha K \ ,
\ee
and ``gamma--driver shift". The gamma--driver condition is usually written as a system of 
two first order (in time) 
equations for $\beta^a$ and an auxiliary field $B^a = 4\dot\beta^a/3$. These equations, along with 
suitable initial conditions, can be integrated to yield a
single first order equation for the shift.\cite{vanMeter:2006vi} Expressed as either two equations or one, 
the gamma--driver condition 
depends on the trace (in its lower indices) of the Christoffel symbols built from the conformal 
metric. In terms of the physical metric, we can write the single--equation form of the gamma--driver 
shift condition as 
\be\label{gammadriver}
	\dot\beta^a - \beta^b\partial_b\beta^a = \frac{3}{4}\sqrt{g}^{2/3} \left[ 
		\Gamma^a_{bc} g^{bc} + \frac{1}{3} g^{ab}\Gamma^c_{bc} \right] - \eta \beta^a \ ,
\ee
where $\eta$ is a constant parameter. 
The first term on the right--hand side (apart from the factor of $3/4$) is the trace of the conformal 
Christoffel symbols. 

Let us see if we can find a set of  functions 
$\hat\Lambda$ and $\hat\Omega^a$ that yield the gauge conditions above, and then ask whether the 
resulting system is strongly hyperbolic. In this example we dispense with any attempt to construct a
formulation that is covariant under spatial diffeomorphisms or time reparametrizations. 
Comparing Eqs.~(\ref{xADMalpha},\ref{xADMbeta}) with 
the 1+log slicing condition 
(\ref{onepluslog}) and the gamma--driver shift condition (\ref{gammadriver}), we find 
\bea
	\hat\Lambda & = & \beta^a\partial_a\alpha - \alpha P/\sqrt{g} + F_1 \pi + F_2^a \rho_a \ ,\\
	\hat\Omega^a & = & \beta^b\partial_b\beta^a + \frac{3}{4}\sqrt{g}^{2/3} \left[ 
		\Gamma^a_{bc} g^{bc} + \frac{1}{3} g^{ab}\Gamma^c_{bc} \right] - \eta \beta^a 
		+ F_3^a\pi + F_4^{ab}\rho_b \ ,
\eea
where the $F$'s are functions of the coordinates $g_{ab}$, $\alpha$, and $\beta^a$ 
(and not their spatial derivatives). In this case the analysis of hyperbolicity is complicated by the 
fact that the scalar and vector blocks of the 
principal symbol are coupled by the terms $F_2^a$, $F_3^a$, and $F_4^{ab}$. The combined 
scalar/vector block has real eigenvalues, but completeness of the eigenvectors can be achieved only if $F_4^{ab}$ 
depends on the normal direction $n_a \equiv k_a/|k|$. This is not acceptable; the Hamiltonian cannot 
depend on the propagation direction of a perturbative solution. We conclude that there is 
no Hamiltonian of the form (\ref{xHamiltonian}) that yields strongly hyperbolic equations with 
1+log slicing (\ref{onepluslog}) and gamma--driver shift (\ref{gammadriver}). 

\section{Concluding Remarks}
This paper outlines a procedure for constructing Hamiltonian formulations of Einstein's theory 
with dynamical gauge conditions and varying levels of hyperbolicity. One can use this procedure 
as a tool to help identify well posed formulations of the evolution equations that also maintain 
Hamiltonian and variational structures. The issues of dynamical gauge conditions and 
hyperbolicity cannot be separated. They are both dictated by the dependence of the  
multipliers $\hat\Lambda$ and $\hat\Omega^a$ on the canonical variables. 
There are many possibilities that one can explore for this dependence.

The Hamiltonian and variational formulations of general relativity have shaped our perspective 
and provided deep insights into the theory. In addition, there are a number of practical uses for a 
Hamiltonian/variational formulation. With an action principle we can pass between 
spacetime and space--plus--time formulations by adding or removing momentum variables. We 
can develop a fully first order multisymplectic version of the theory. We can also 
develop new computational techniques such as variational and symplectic 
integrators.\cite{Marsden:2001,Brown:varint} 

One important issue that has not been addressed here is the constraint evolution system. In 
numerical simulations it is important to control the growth of constraint violations. This 
might be accomplished in the present framework by including appropriate terms in $\hat\Lambda$ 
and $\hat\Omega^a$ to ensure that the constraints are damped. For example, a damping 
term $-C\pi$ (where $C$ is a positive constant) can be added to the $\dot\pi$ equation by 
including a lower order term $C\alpha$ in $\hat\Lambda$. This issue will be explored in 
more detail elsewhere.\cite{BrownPrep}

The formalism outlined here can be further extend by introducing 
dynamical equations for $\Lambda$ and 
$\Omega^a$. This is accomplished by introducing momentum variables conjugate to these multipliers. 
The new momentum variables are primary constraints and are accompanied    
by a new set of undetermined multipliers. Dynamical equations for $\Lambda$ and $\Omega^a$ 
are introduced by allowing the new multipliers to depend on the canonical variables. In this way 
we can construct gauge driver conditions similar to the ones used with 
the generalized harmonic formulation.\cite{Pretorius:2004jg,Lindblom:2007xw} We can also 
allow for gauge conditions that are expressed as systems of PDE's, such as the two--equation 
version of the gamma--driver shift condition. 

\section{Acknowledgments}
I would like to thank Claudio Bunster for his guidance and support and, most of all, 
for his inspiration through the years. 
I also thank Lee Lindblom, Mark Scheel, and Manuel Tiglio for discussions 
that helped stimulate the ideas presented here. 
I especially thank Olivier Sarbach for his patient explanations of hyperbolicity. 
This work was supported by NSF Grant PHY-0600402.


\end{document}